\documentclass[11pt,showpacs,preprint,preprintnumbers,nofootinbib,
superscriptaddress,amsmath,amssymb]{revtex4}

\usepackage{graphicx}
\usepackage{dcolumn}
\usepackage{bm}
\usepackage{amssymb}
\usepackage{amsmath}
\usepackage{epsfig}    
\usepackage{color}
\usepackage{slashed}

\begin{document} 

\title{Radiative corrections to the Yukawa coupling constants \\
in two Higgs doublet models }

\preprint{UT-HET 086}
\author{Shinya Kanemura}
\affiliation{Department of Physics, University of Toyama, \\3190 Gofuku, Toyama 930-8555, Japan}
\author{Mariko Kikuchi}
\affiliation{Department of Physics, University of Toyama, \\3190 Gofuku, Toyama 930-8555, Japan}
\author{Kei Yagyu}
\affiliation{Department of Physics, National Central University, \\Chungli 32001, Taiwan}

\begin{abstract}

We calculate one-loop corrected Yukawa coupling constants $hf\bar{f}$ 
for the standard model like Higgs boson $h$ 
in two Higgs doublet models. 
We focus on the models with the softly-broken $Z_2$ symmetry, which is imposed to avoid the flavor changing neutral current. 
Under the $Z_2$ symmetry, there are four types of Yukawa interactions. 
We find that one-loop contributions from extra Higgs bosons modify the $hf\bar{f}$ couplings to be maximally about $5\%$ under 
the constraint from perturbative unitarity and vacuum stability.
Our results show that the pattern of tree-level deviations by the mixing effect in each type of 
Yukawa couplings from the SM predictions does not change 
even including radiative corrections.  
Moreover, when the gauge couplings $hVV$ ($V=W,Z$) are found to be slightly (with a percent level) differ from the 
SM predictions, the $hf\bar{f}$ couplings also deviate but more largely. 
Therefore, in such a case, not only can we determine the type of 
Yukawa couplings but also we can obtain information on the 
extra Higgs bosons by comparing the predictions with  
precisely measured $hf\bar{f}$ and $hVV$ couplings at future electron-positron colliders.

\end{abstract}
\maketitle

\section{Introduction}

By the discovery of a Higgs boson at the CERN Large Hadron Collider (LHC)~\cite{LHC_Higgs}, 
the standard model (SM) has been completed. 
So far, within the error properties of the observed boson are 
consistent with those of the Higgs boson in the SM such as the mass, the CP parity and the signal strengths.
Thus, the discovered boson can be regarded as the SM-like Higgs boson $h$.  

However, this fact does not necessarily mean that the SM is correct in a fundamental level, because 
the SM-like Higgs boson can be described not only 
in the minimal Higgs sector with only one isospin scalar doublet but also in non-minimal Higgs sectors.  
In fact, the minimal Higgs sector of the SM is nothing but an assumption without any principle. 
In addition, non-minimal Higgs sectors often appear in physics models beyond the SM in which 
several unsolved problems such as   
neutrino oscillation, the existence of dark matter and the baryon asymmetry of the Universe 
within the SM are tried to be explained. 
Therefore, non-minimal Higgs sectors (e.g., with additional singlets, doublets and/or triplets) 
should be comprehensively studied to determine the true structure of the Higgs sector and 
to probe new physics models. 

In extended Higgs sectors, the Higgs boson coupling constants can be deviated from the corresponding SM predictions. 
In addition, a pattern of the deviations strongly depends on properties of the Higgs sector; i.e., the number of Higgs fields and their quantum numbers. 
Therefore, by $``Fingerprinting"$, i.e., by comparing the deviations in various Higgs boson couplings with the theory predictions, 
we can extract the structure of the Higgs sector. 
 
The Higgs boson couplings will be measured at future colliders as precisely as possible. 
For example, the $hVV$ ($V=W,~Z$) and $hf\bar{f}$ ($f=t,~b,~\tau$) couplings are supposed to be measured with 
approximately $5\%$ and $10\%$ accuracies at the LHC with the collision energy 
to be 14 TeV and the integrated luminosity to be 300~fb$^{-1}$, respectively~\cite{Peskin,ILC_white, ILC_TDR}. 
Moreover, they are expected to be measured with typically $1\%$ at the International Linear Collider (ILC)
with the collision energy to be $500$ GeV and the integrated luminosity to be $500$ fb$^{-1}$~\cite{Peskin,ILC_TDR,Plehn,ILC_white}. 

In this Letter, we calculate deviations in the Yukawa couplings from the SM predictions in two Higgs doublet models (THDMs) 
at the one-loop level, especially focusing on those for the SM-like Higgs boson $h$. 

THDMs are a simple but well-motivated example for extended Higgs sectors. 
First, the electroweak rho parameter is naturally predicted to be unity at the tree level, whose experimental value 
is close to unity; i.e., $\rho_{\text{exp}}=1.0004^{+0.0003}_{-0.0004}$~\cite{PDG}.  
In the other extended Higgs sectors such as Higgs triplet models, 
the rho parameter is not guaranteed to be unity at the tree level.  
Although even in the THDMs the rho parameter can deviate from unity due to the one-loop correction~\cite{rho_loop}, 
its amount can easy be within the error of the measurement\footnote{
One-loop corrections to the rho parameter in Higgs triplet models have been discussed in Refs.~\cite{rho_triplet}.}. 
Second, the Higgs sector in several new physics models has the structure of the THDM. 
For example, the supersymmetry requires at least two Higgs doublets. 
Neutrino mass models such as 
radiative seesaw models~\cite{zee,ma,aks,t7} 
and the neutrinophilic model~\cite{nuphilic}
contain two Higgs doublet fields in their Higgs sector. 
The hierarchy between top and bottom quark masses may be naturally explained in the THDM~\cite{Hashimoto}. 
Furthermore, additional CP violating phases can appear, and the strong first order electroweak phase transition can 
occur due to nondecoupling effects of extra scalar bosons. 
These characteristics are required to realize the successful electroweak baryogenesis scenario~\cite{EWBG}.
A comprehensive review of various classes of the THDM is given in Ref.~\cite{2HDM_rev}. 

Unlike the SM, in multi-doublet models, the mass matrix for fermions and the interaction matrix 
among a neutral Higgs boson and fermions cannot be diagonalized simultaneously. 
That causes flavor changing neutral currents (FCNCs) at the tree level, 
which are severely constrained from flavor experiments
such as $K_L^0\to \mu^+\mu^-$, $B^0$-$\bar{B}^0$ mixing and so on.
In order to avoid the tree level FCNC, a discrete $Z_2$ symmetry~\cite{GW} may be imposed as the simplest way.  
If we consider the case with the softly-broken $Z_2$ symmetry\footnote{
The unbroken, even by the vacuum, $Z_2$ symmetric THDM is known as the inert doublet model~\cite{IDM}. }, 
there are four independent types of Yukawa interactions under the different charge assignments to quarks and 
charged leptons~\cite{4types,Akeroyd}. 
We call them Type-I, Type-II, Type-X and Type-Y THDMs~\cite{typeX}. 
A lot of phenomenological studies in these THDMs have been performed before the Higgs boson discovery~\cite{THDMs_pheno_before} and 
after that~\cite{THDMs_pheno_after}. 
Each type of THDMs can be related to various new physics models. 
For example, the Higgs sector in the minimal supersymmetric SM (MSSM) corresponds to the Type-II THDM with supersymmetric relations. 
On the other hand, the Type-X THDM is applied to radiative seesaw models~\cite{aks,t7}. 
Therefore, discrimination of the types of Yukawa interactions in the THDM 
is important to test new physics models. 

In order to compare precisely measured Higgs boson couplings as mentioned above, 
we need to prepare precise calculations of the Higgs boson couplings in various Higgs sectors.  
Namely, it is essentially important to take into account the effects of radiative corrections. 
So far, there are several studies of one-loop calculations for the Higgs boson couplings in various versions of the THDM. 
One-loop corrections to the triple Higgs boson coupling $hhh$~\cite{Hollik_hhh} and Yukawa couplings~\cite{Hollik_Yukawa}
have been calculated in the MSSM~Higgs sector. 
In the softly-broken $Z_2$ symmetric THDM, 
the $hhh$ and $hVV$ couplings have also been calculated at the one-loop level in Ref.~\cite{KOSY}. 
However, one-loop corrected Yukawa couplings have not been systematically analysed in the four types of THDMs. 
In this Letter, we would like to clarify how the tree level deviations in various Yukawa couplings 
shown in Ref.~\cite{ILC_white} can be modified by the one-loop corrections.  

\section{Two Higgs doublet models}

\begin{table}[t]
\begin{center}
{\renewcommand\arraystretch{1.2}
\begin{tabular}{c|ccccccc|ccccccccc}\hline\hline
&
\multicolumn{7}{c|}{$Z_2$ charge}
&\multicolumn{9}{c}{Mixing factor}\\\cline{2-17}
&$\Phi_1$&$\Phi_2$&$Q_L$&$L_L$&
$u_R$&$d_R$&$e_R$
&$\xi_h^u$ &$\xi_h^d$&$\xi_h^e$&$\xi_H^u$&$\xi_H^d$&$\xi_H^e$ &$\xi_A^u$&$\xi_A^d$&$\xi_A^e$\\\hline
Type-I &$+$&
$-$&$+$&$+$&
$-$&$-$&$-$&
$\frac{\cos\alpha}{\sin\beta}$&$\frac{\cos\alpha}{\sin\beta}$&$\frac{\cos\alpha}{\sin\beta}$&$\frac{\sin\alpha}{\sin\beta}$&$\frac{\sin\alpha}{\sin\beta}$&$\frac{\sin\alpha}{\sin\beta}$&$\cot\beta$&$-\cot\beta$&$-\cot\beta$\\\hline
Type-II&$+$&
$-$&$+$&$+$&
$-$
&$+$&$+$
&$\frac{\cos\alpha}{\sin\beta}$&$-\frac{\sin\alpha}{\cos\beta}$&$-\frac{\sin\alpha}{\cos\beta}$&$\frac{\sin\alpha}{\sin\beta}$&$\frac{\cos\alpha}{\cos\beta}$&$\frac{\cos\alpha}{\cos\beta}$&$\cot\beta$&$\tan\beta$&$\tan\beta$\\\hline
Type-X &$+$&
$-$&$+$&$+$&
$-$
&$-$&$+$
&$\frac{\cos\alpha}{\sin\beta}$&$\frac{\cos\alpha}{\sin\beta}$&$-\frac{\sin\alpha}{\cos\beta}$&$\frac{\sin\alpha}{\sin\beta}$&$\frac{\sin\alpha}{\sin\beta}$&$\frac{\cos\alpha}{\cos\beta}$&$\cot\beta$&$-\cot\beta$&$\tan\beta$\\\hline
Type-Y &$+$&
$-$&$+$&$+$&
$-$
&$+$&$-$
&$\frac{\cos\alpha}{\sin\beta}$&$-\frac{\sin\alpha}{\cos\beta}$&$\frac{\cos\alpha}{\sin\beta}$&$\frac{\sin\alpha}{\sin\beta}$&$\frac{\cos\alpha}{\cos\beta}$&$\frac{\sin\alpha}{\sin\beta}$&$\cot\beta$&$\tan\beta$&$-\cot\beta$\\\hline\hline
\end{tabular}}
\caption{Charge assignment of the softly broken $Z_2$ symmetry and the mixing factors in Yukawa interactions given in Eq.~(\ref{yukawa_thdm})~\cite{typeX}.}
\label{yukawa_tab}
\end{center}
\end{table}
 
In this Letter, we assume the CP-conservation of the Higgs sector. 
Let us fix the $Z_2$ charge
for the two Higgs doublet fields $\Phi_1$ and $\Phi_2$ and 
the left-handed lepton-doublet and quark-doublet fields $L_L$ and $Q_L$
as $+$, $-$, $+$ and $+$, respectively. 
In this set up, four types of the Yukawa interactions are defined by the choice of the $Z_2$ charge assignment for right-handed 
up-type quarks $u_R$, down-type quarks $d_R$ and charged leptons $e_R$ as listed in Table~\ref{yukawa_tab}. 

The Yukawa Lagrangian is then given by
\begin{align}
{\mathcal L}^Y_\text{THDM} =
&-Y_{u}{\overline Q}_Li\sigma_2\Phi^*_uu_R^{}
-Y_{d}{\overline Q}_L\Phi_dd_R^{}
-Y_{e}{\overline L}_L\Phi_e e_R^{}+\text{h.c.},
\end{align}
where $\Phi_{u,d,e}$ are $\Phi_1$ or $\Phi_2$. 
The two doublet fields can be parameterized as 
\begin{align}
\Phi_i=\left[\begin{array}{c}
w_i^+\\
\frac{1}{\sqrt{2}}(v_i+h_i+iz_i)
\end{array}\right],\hspace{3mm}(i=1,2), 
\end{align}
where $v_1$ and $v_2$ are the vacuum expectation values (VEVs) for $\Phi_1$ and $\Phi_2$, 
which satisfy $v\equiv\sqrt{v_1^2+v_2^2}=(\sqrt{2}G_F)^{-1/2}$. 
The ratio of the two VEVs is defined as $\tan\beta=v_2/v_1$.  

The mass eigenstates for the scalar bosons are obtained by the following orthogonal transformations as
\begin{align}
\left(\begin{array}{c}
w_1^\pm\\
w_2^\pm
\end{array}\right)&=R(\beta)
\left(\begin{array}{c}
G^\pm\\
H^\pm
\end{array}\right),\quad 
\left(\begin{array}{c}
z_1\\
z_2
\end{array}\right)
=R(\beta)\left(\begin{array}{c}
G^0\\
A
\end{array}\right),\quad
\left(\begin{array}{c}
h_1\\
h_2
\end{array}\right)=R(\alpha)
\left(\begin{array}{c}
H\\
h
\end{array}\right), \notag\\
\text{with}~R(\theta) &= 
\left(
\begin{array}{cc}
\cos\theta & -\sin\theta\\
\sin\theta & \cos\theta
\end{array}\right),
\label{mixing}
\end{align}
where $G^\pm$ and $G^0$ are the Nambu-Goldstone bosons absorbed by the longitudinal component of $W^\pm$ and $Z$, respectively.  
As the physical degrees of freedom, 
we have a pair of singly-charged Higgs boson $H^\pm$, a CP-odd Higgs boson $A$ and two CP-even Higgs bosons $h$ and $H$. 
We define $h$ as the SM-like Higgs boson with the mass of about 126 GeV. 

The Higgs potential under the softly broken $Z_2$ symmetry and the CP invariance is given by  
\begin{align}
V_\text{THDM} &=m_1^2|\Phi_1|^2+m_2^2|\Phi_2|^2-m_3^2(\Phi_1^\dagger \Phi_2 +\text{h.c.})\notag\\
& +\frac{1}{2}\lambda_1|\Phi_1|^4+\frac{1}{2}\lambda_2|\Phi_2|^4+\lambda_3|\Phi_1|^2|\Phi_2|^2+\lambda_4|\Phi_1^\dagger\Phi_2|^2
+\frac{1}{2}\lambda_5\left[(\Phi_1^\dagger\Phi_2)^2+\text{h.c.}\right]. \label{pot_thdm2}
\end{align}
Eight parameters in the potential are translated into eight physical parameters; namely, 
the masses of $h$, $H$, $A$ and $H^\pm$, two mixing angles $\alpha$ and $\beta$ appearing in Eq.~(\ref{mixing}), 
the VEV $v$ and the remaining parameter $M^2$ defined by 
\begin{align}
M^2=\frac{m_3^2}{\sin\beta\cos\beta}, \label{bigm}
\end{align}
which describes the soft breaking scale of the $Z_2$ symmetry. 
Exact formulae for the Higgs boson masses and the mixing angle $\alpha$ are given in Ref.~\cite{KOSY}. 

The Yukawa interactions are expressed in terms of mass eigenstates
of the Higgs bosons as
\begin{align}
{\mathcal L}^Y_\text{THDM} 
&=
-\sum_{f=u,d,e}\frac{m_f}{v} \left( \xi_h^f{\overline
f}fh+\xi_H^f{\overline
f}fH-i\xi_A^f{\overline f}\gamma_5fA\right)\notag\\
&+\left[\frac{\sqrt2V_{ud}}{v}\overline{u}
\left(m_u\xi_A^u\text{P}_L+m_d\xi_A^d\text{P}_R\right)d\,H^+
+\frac{\sqrt2m_\ell\xi_A^e}{v}\overline{\nu^{}}P_Re^{}H^+
+\text{h.c.}\right],\label{yukawa_thdm}
\end{align}
where the factors $\xi^f_\varphi$ are listed in Table~\ref{yukawa_tab}.

We here summarize the tree level scale factors of $h$ for the $hVV$ ($V=W,Z$) and $hf\bar{f}$ couplings, 
which are defined by the value of the coupling constants divided by the corresponding SM values as follows
\begin{align}
&\kappa_V = \sin(\beta-\alpha)\equiv \sqrt{1-\delta}~(0\leq\delta\leq 1)~\text{for all types},\label{kv}\\
&\kappa_u=\xi_h^u\simeq 1+\vartheta \cot\beta \sqrt{\delta}-\frac{\delta}{2}~\text{for all types},\label{ku}\\
&\kappa_d=\xi_h^d\simeq 1+\vartheta \cot\beta \sqrt{\delta}-\frac{\delta}{2}~~~\left(1-\vartheta \tan\beta \sqrt{\delta}-\frac{\delta}{2}\right)~\text{for Type-I,-X (Type-II,-Y)},\label{kd}\\
&\kappa_e=\xi_h^e\simeq 1+\vartheta \cot\beta \sqrt{\delta}-\frac{\delta}{2}~~~\left(1-\vartheta \tan\beta \sqrt{\delta}-\frac{\delta}{2}\right)~\text{for Type-I,-Y (Type-II,-X)}, \label{ke}
\end{align}
where $\delta$ and $\vartheta$ are $\cos^2(\beta-\alpha)$ and the sign of $\cos(\beta-\alpha)$, respectively, in the THDMs.  
The nearly-equals in $\kappa_f$ are valid in the case of $\delta\ll 1$. 
Clearly, when $\sin(\beta-\alpha)=1$ (or equivalently taking $\delta=0$) is taken, all the scale factors given in 
Eqs.~(\ref{kv})-(\ref{ke}) become unity, which mean all the tree level $hVV$ and $hf\bar{f}$ couplings are getting the same value 
as in the SM. 
We then define the SM-like limit by $\sin(\beta-\alpha)\to 1$. 
The other Higgs bosons; namely $H^\pm$, $A$ and $H$, should be regarded as extra Higgs bosons. 
As long as we discuss in the SM-like region, 
the squared masses of the extra Higgs bosons are given by the following form
\begin{align}
m_{\Phi}^2 =\lambda_iv^2+M^2,\quad \Phi=H^\pm,~A,~H, \label{mphi}
\end{align}
where $\lambda_i$ represent some combinations of the $\lambda$ couplings given in Eq.~(\ref{pot_thdm2}).  
We note that in general, the mass formula for $H$ is rather complicated than Eq.~(\ref{mphi}). 
However, when we take $\sin(\beta-\alpha)=1$, the expression in Eq.~(\ref{mphi}) also holds for $H$. 
See Ref.~\cite{KOSY} for the explicit formula.

\section{Renormalization}

\begin{table}[t]
\begin{center}
{\renewcommand\arraystretch{1.3}
\begin{tabular}{c||ccc}\hline\hline
&$\delta\xi_h^u$ &$\delta\xi_h^d$&$\delta\xi_h^e$\\\hline
Type-I  &$-\frac{\cos\alpha}{\sin\beta}(\cot\beta\delta \beta+\tan\alpha\delta\alpha )$
&$-\frac{\cos\alpha}{\sin\beta}(\cot\beta\delta \beta+\tan\alpha\delta\alpha)$&$-\frac{\cos\alpha}{\sin\beta}(\cot\beta\delta \beta+\tan\alpha\delta\alpha )$ \\\hline
Type-II &$-\frac{\cos\alpha}{\sin\beta}(\cot\beta\delta \beta+\tan\alpha\delta\alpha )$&
$-\frac{\sin\alpha}{\cos\beta}(\tan\beta \delta\beta+\cot\alpha \delta \alpha)$
&$-\frac{\sin\alpha}{\cos\beta}(\tan\beta \delta\beta+\cot\alpha \delta \alpha)$
\\\hline
Type-X  &$-\frac{\cos\alpha}{\sin\beta}(\cot\beta\delta \beta+\tan\alpha\delta\alpha )$&$-\frac{\cos\alpha}{\sin\beta}(\cot\beta\delta \beta+\tan\alpha\delta\alpha )$&$-\frac{\sin\alpha}{\cos\beta}(\tan\beta \delta\beta+\cot\alpha \delta \alpha)$\\\hline
Type-Y  &$-\frac{\cos\alpha}{\sin\beta}(\cot\beta\delta \beta+\tan\alpha\delta\alpha )$&$-\frac{\sin\alpha}{\cos\beta}(\tan\beta \delta\beta+\cot\alpha \delta \alpha)$
&$-\frac{\cos\alpha}{\sin\beta}(\cot\beta\delta \beta+\tan\alpha\delta\alpha )$\\\hline\hline
\end{tabular}}
\caption{The counter term for the mixing factors in Yukawa interactions.}
\label{delxi}
\end{center}
\end{table}

In this section, we calculate one-loop corrected Yukawa couplings for the SM-like Higgs boson $h$ 
in the four types of Yukawa interactions based on the on-shell renormalization scheme. 
For the calculation of each diagram, we choose the 't Hooft-Feynman gauge. 
The renormalized $hf\bar{f}$ vertex can be expressed by the following three parts, 
\begin{align}
\hat{\Gamma}_{hff}(p_1^2,p_2^2,q^2)=\Gamma_{hff}^{\text{tree}}+\delta \Gamma_{hff}+\Gamma_{hff}^{\text{1PI}}(p_1^2,p_2^2,q^2), \label{ryuk}
\end{align}
where $p_1^\mu$ and $p_2^\mu$ are the incoming momenta for the fermion and anti-fermion, and $q^\mu~(=p_1^\mu+p_2^\mu)$ 
is the outgoing momentum for $h$. 
In Eq.~(\ref{ryuk}), the first, second and third terms in the right hand side are 
the contributions from 
the tree level diagram, the counter terms and the 1PI diagrams to the $hf\bar{f}$ couplings, respectively. 
The tree level contribution is obtained in terms of the mixing factor listed in Table~\ref{yukawa_tab}. 

The counter term contribution is given by
\begin{align}
\delta \Gamma_{hff}=-i\frac{m_f}{v}\xi_h^f\left[\frac{\delta m_f}{m_f}+\delta Z_V^f+\frac{1}{2}\delta Z_h
+\frac{\delta\xi_h^f}{\xi_h^f}+\frac{\xi_H^f}{\xi_h^f}(\delta C_h+\delta \alpha)-\frac{\delta v}{v}\right], \label{cyuk}
\end{align}
where $\delta\xi_h^f$ depend on the type of Yukawa interaction, which are listed in Table~\ref{delxi}. 
In the following, we explain how each of the counter terms in Eq.~(\ref{cyuk}) can be determined. 
The counter terms for the fermion mass and the wave function renormalization are given by
\begin{align}
m_f &\to m_f+\delta m_f,\quad
\psi_L \to \left(1+\frac{1}{2}\delta Z_L^f\right)\psi_L,\quad 
\psi_R \to \left(1+\frac{1}{2}\delta Z_R^f\right)\psi_R, \label{del_f}
\end{align}
where $\psi_L$ and $\psi_R$ are the left-handed and right-handed fermions. 
The renormalized fermion two point function is expressed by the following two parts; 
\begin{align}
\hat{\Pi}_{ff}(p^2) &= \hat{\Pi}_{ff,V}(p^2) +\hat{\Pi}_{ff,A}(p^2), 
\end{align}
where
\begin{align}
\hat{\Pi}_{ff,V}(p^2) &= 
p\hspace{-2mm}/\left[\Pi_{ff,V}^{\text{1PI}}(p^2) + \delta Z_V^f\right] 
+m_f \left[\Pi_{ff,S}^{\text{1PI}}(p^2)-\delta Z_V^f-\frac{\delta m_f}{m_f}\right], \notag\\
\hat{\Pi}_{ff,A}(p^2) &= - p\hspace{-2mm}/\gamma_5\left[\Pi_{ff,A}^{\text{1PI}}(p^2) + \delta Z_A^f\right],  \label{fermi_2p}
\end{align}
with
\begin{align}
&\delta Z_V^f = \frac{\delta Z_L^f+\delta Z_R^f}{2},\quad  \delta Z_A^f = \frac{\delta Z_L^f -\delta Z_R^f}{2}. 
\end{align}
In Eq.~(\ref{fermi_2p}), $\Pi_{ff,V}^{\text{1PI}}$, $\Pi_{ff,A}^{\text{1PI}}$ and $\Pi_{ff,S}^{\text{1PI}}$ 
are the vector, axial vector and scalar parts of the 1PI diagram contributions at the one-loop level, respectively. 
By imposing the three renormalization conditions
\begin{align}
&\hat{\Pi}_{ff,V}(m_f^2) = 0, \notag\\
&\frac{d}{dp\hspace{-2mm}/}\hat{\Pi}_{ff,V}(p^2)\Big|_{p^2 =m_f^2} = 0, \quad 
\frac{d}{dp\hspace{-2mm}/}\hat{\Pi}_{ff,A}(p^2)\Big|_{p^2 =m_f^2} = 0,
\label{rc_f2}
\end{align}
we obtain 
\begin{align}
\frac{\delta m_f}{m_f} &= \Pi_{ff,V}^{\text{1PI}}(m_f^2) +\Pi_{ff,S}^{\text{1PI}}(m_f^2),\notag\\
\delta Z_V^f
&=-\Pi_{ff,V}^{\text{1PI}}(m_f^2) -2m_f^2\left[\frac{d}{dp^2}\Pi_{ff,V}^{\text{1PI}}(p^2)\Big|_{p^2=m_f^2}+\frac{d}{dp^2}\Pi_{ff,S}^{\text{1PI}}(p^2)\Big|_{p^2=m_f^2}\right], \notag\\
\delta Z_A^f &= - \Pi_{ff,A}^{\text{1PI}}(m_f^2)+2m_f^2\frac{d}{dp^2}\Pi_{ff,A}^{\text{1PI}}(p^2)\Big|_{p^2=m_f^2}. 
\label{del_zvf}
\end{align}
Although the counter term $\delta Z_A^f$ is not used in the following discussion, we here show the expression for completeness. 

According to Ref.~\cite{KOSY}, the counter terms $\delta Z_h$, $\delta C_h$ and $\delta \alpha$ are defined in the CP-even Higgs sector as
\begin{align}
&\left(\begin{array}{c}
H\\
h
\end{array}\right) 
\to \left(\begin{array}{cc}
1+\frac{1}{2}\delta Z_H & \delta C_h+\delta\alpha\\
\delta C_h -\delta\alpha & 1+\frac{1}{2}\delta Z_h
\end{array}\right)\left(\begin{array}{c}
H\\
h
\end{array}\right). 
\end{align}
In order to determine them, 
we impose the on-shell conditions for the scalar two point functions; 
\begin{align}
&\frac{d}{dp^2}\hat{\Pi}_{hh}(p^2)\Big|_{p^2=m_h^2}=0,  \quad
\hat{\Pi}_{Hh}(p^2=m_H^2)=\hat{\Pi}_{Hh}(p^2=m_h^2)=0,\label{scalar_rc}
\end{align}
where $\hat{\Pi}_{hh}$ and $\hat{\Pi}_{Hh}$ are the renormalized two point functions of $hh$ and $Hh$. 
From the three conditions given in Eq.~(\ref{scalar_rc}), 
three counter terms $\delta Z_h$, $\delta\alpha$ and $\delta C_h$ are determined.

The counter term $\delta\beta$, which is defined by the shift $\beta\to \beta+\delta\beta$, is determined by 
requiring that the mixing between $A$ and $G^0$ is absent at the on-shell for $A$ and $G^0$.
This can be expressed in terms of the renormalized $A$-$G^0$ mixing $\hat{\Pi}_{AG}$ as 
\begin{align}
\hat{\Pi}_{AG}(p^2=m_Z^2)=\hat{\Pi}_{AG}(p^2=m_A^2)=0. \label{GA}
\end{align}
In fact, we can determine not only $\delta \beta$ but also the counter term associated with the mixing between the CP-odd states 
$\delta C_A$ corresponding to $\delta C_h$ in the CP-even sector. 

We here note that the condition given in Eq.~(\ref{GA}) with $p^2=m_A^2$ 
is equivalent to the requirement for the vanishing $Z$-$A$ mixing due to 
the Ward-Takahashi identity; i.e., 
\begin{align}
\hat{\Pi}_{ZA}(p^2=m_A^2)=0,\label{ZA}
\end{align}
where $\hat{\Pi}_{ZA}$ is defined from the renormalized $Z$-$A$ mixing $\hat{\Pi}_{ZA}^\mu=-ip^\mu\hat{\Pi}_{ZA}$. 
According to Ref.~\cite{gauge_depend}, the determination of $\delta \beta$ by Eq.~(\ref{GA}) or (\ref{ZA}) has a gauge dependence of order
$m_Z^2/m_A^2$. We neglect such a dependence in the following discussion, because it is not essentially important in our numerical results. 

The counter term for the VEV $\delta v$ is determined from the renormalization of the electroweak parameters. 
We determine the counter terms for the masses of $W$ and $Z$ bosons 
and the fine structure constant according to the electroweak on-shell scheme~\cite{Hollik}, 
so that we obtain 
\begin{align}
\frac{\delta v}{v}&= \frac{1}{2}
\left[
\frac{s_W^2-c_W^2}{s_W^2}\frac{\Pi_{WW}^{\text{1PI}}(m_W^2)}{m_W^2}+\frac{c_W^2}{s_W^2}\frac{\Pi_{ZZ}^{\text{1PI}}(m_Z^2)}{m_Z^2}
-\frac{d}{dp^2}\Pi_{\gamma\gamma}^{\text{1PI}}(p^2)\Big|_{p^2=0}+\frac{2s_W}{c_W}\frac{\Pi_{Z\gamma}^{\text{1PI}}(0)}{m_Z^2}\right], \label{del_vev} 
\end{align}
where $\Pi_{XY}^{\text{1PI}}$ are the contributions from the 1PI diagrams for the gauge boson self-energies 
and $c_W=\cos\theta_W$ and $s_W=\sin\theta_W$ with $\theta_W$ being the weak mixing angle. 
Instead of the calculation of $\frac{d}{dp^2}\Pi_{\gamma\gamma}^{\text{1PI}}(p^2)\Big|_{p^2=0}$, we introduce 
the shift of the fine structure constant $\alpha_{\text{em}}$ from $0$ to the scale of $m_Z$ as 
\begin{align}
\Delta \alpha_{\text{em}}= \frac{d}{dp^2}\Pi_{\gamma\gamma}^{\text{1PI}}(p^2)\Big|_{p^2=0}
-\frac{d}{dp^2}\Pi_{\gamma\gamma}^{\text{1PI}}(p^2)\Big|_{p^2=m_Z^2}.\label{Del_alpha}
\end{align}

Finally, the 1PI contributions to the $hf\bar{f}$ vertex can be decomposed into the following 
eight form factors in general;
\begin{align}
&\Gamma_{hff}^{\text{1PI}}(p_1^2,p_2^2,q^2) = \notag\\
&F_{hff}^S+\gamma_5 F_{hff}^P+p_1\hspace{-3.5mm}/\hspace{2mm}F_{hff}^{V1}
+p_2\hspace{-3.5mm}/\hspace{2mm}F_{hff}^{V2}+p_1\hspace{-3.5mm}/\hspace{2mm}\gamma_5 F_{hff}^{A1}
+p_2\hspace{-3.5mm}/\hspace{2mm}\gamma_5F_{hff}^{A2}
+p_1\hspace{-3.5mm}/\hspace{2mm}p_2\hspace{-3.5mm}/\hspace{2mm}F_{hff}^{T}
+p_1\hspace{-3.5mm}/\hspace{2mm}p_2\hspace{-3.5mm}/\hspace{2mm}\gamma_5F_{hff}^{PT}. 
\end{align}
We note that in the on-shell case; i.e., $p_1^2=p_2^2=m_f^2$, 
the form factors proportional to $\gamma_5$ are vanished in the SM-like limit, so that 
only $F_{hff}^S$, $F_{hff}^{V1}$, $F_{hff}^{V2}$ and $F_{hff}^{T}$ are survived. 
Among those form factors, $F_{hff}^S$ gives the dominant contribution to the $hf\bar{f}$ vertex. 

We show the expression of the deviation in renormalized Yukawa coupling from the SM prediction. 
Because the general expression is rather complicated, 
we here give the formula in the case of $\sin(\beta-\alpha)=1$ and $m_{H^+}=m_A=m_H~(\equiv m_\Phi)$ 
in terms of the Passarino-Veltman functions~\cite{PV}; 
\begin{align}
&\hat{\Gamma}_{hff}^{\text{THDM}}(m_f^2,m_f^2,m_h^2)\notag\\
&\simeq 
\hat{\Gamma}_{hff}^{\text{SM}}(m_f^2,m_f^2,m_h^2)
+\frac{m_f}{v}\frac{1}{16\pi^2}\Bigg\{
\frac{2m_{f'}^2}{v^2}\xi_A^{d}\cot\beta\Big[(m_h^2-2m_f^2)C_{12}(m_{f'},m_\Phi,m_{f'})\notag\\
&+(2m_{f'}^2-m_f^2)C_0(m_{f'},m_\Phi,m_{f'})+v\lambda_{\Phi\Phi h}C_0(m_{\Phi},m_{f'},m_{\Phi}) \Big]
\notag\\
&+4\lambda_{\Phi\Phi h}^2\frac{d}{dp^2}B_0(p^2;m_{\Phi},m_{\Phi})\Big|_{p^2=m_h^2}
-\frac{6m_t^2}{v^2} I_f\xi_A^f\cot\beta B_0(m_\Phi^2;m_t,m_t)\notag\\
&+\frac{6m_t^4}{v^2(m_\Phi^2-m_h^2)}
I_f\xi_A^f\cot\beta\left[\left(4-\frac{m_h^2}{m_t^2}\right)B_0(m_h^2;m_t,m_t)-\left(4-\frac{m_\Phi^2}{m_t^2}\right)B_0(m_\Phi^2;m_t,m_t) \right]\notag\\
&+\frac{6\lambda_{\Phi\Phi h}\lambda_{\Phi\Phi H}}{m_\Phi^2-m_h^2}I_f\xi_A^f
\left[B_0(m_h^2;m_\Phi,m_\Phi)-B_0(m_\Phi^2;m_\Phi,m_\Phi) \right]
\Bigg\},
\label{limit1}
\end{align}
where $f'$ is the fermion whose electromagnetic charge is different by one unit from $f$, and 
$I_f=+1/2~(-1/2)$ for $f=u$ ($d,e$). 
The scalar three-point couplings are given by 
\begin{align}
\lambda_{\Phi\Phi h}=\frac{m_h^2+2m_\Phi^2-2M^2}{v},\quad 
\lambda_{\Phi\Phi H}=\frac{M^2-m_\Phi^2}{v}\cot2\beta. \label{limit2}
\end{align}
The shortened notations are used such as 
$C_i(m_f^2,m_f^2,m_h^2;m_1,m_2,m_3)=C_i(m_1,m_2,m_3)$ in Eq.~(\ref{limit1}). 
We will give the full one-loop expression in the general case elsewhere~\cite{full}.

\section{Results}

In this section, we show the numerical results. 
We use the following inputs~\cite{PDG}; 
\begin{align}
&m_Z=91.1875~\text{GeV},~G_F=1.16639\times 10^{-5}~\text{GeV}^{-2},~\alpha_{\text{em}}^{-1}=137.035989,~\Delta\alpha_{\text{em}}=0.06635, \notag\\
&m_t=173.07~\text{GeV},~m_b=4.66~\text{GeV},~m_c=1.275~\text{GeV},~m_\tau=1.77684~\text{GeV}. 
\end{align}
We here take all the extra Higgs boson masses to be the same; i.e., 
$m_{H^+}=m_A=m_H~(= m_\Phi)$ for avoiding the constraint from the electroweak rho parameter~\cite{rho_loop}. 
In the THDM, theoretical bounds from perturbative unitarity and vacuum stability have been derived
in Refs.~\cite{pv_THDM} and \cite{vs_THDM}, respectively, 
and we take into account them using formulae given in Ref.~\cite{KOSY}. 
We will show more comprehensive choice of parameters elsewhere~\cite{full}.

We evaluate the one-loop renormalized scale factors defined by
\begin{align}
\hat{\kappa}_f \equiv
\frac{
\hat{\Gamma}_{hff}(m_f^2,m_f^2,m_h^2)_{\text{THDM}}
}{\hat{\Gamma}_{hff}(m_f^2,m_f^2,m_h^2)_{\text{SM}}}, \quad \text{for}~f=c,~b,~\tau, \label{kappa_f}
\end{align}
where 
\begin{align}
\hat{\Gamma}_{hff}(m_f^2,m_f^2,m_h^2)= \Gamma_{hff}^{\text{tree}}+\delta \Gamma_{hff}+F_S(m_f^2,m_f^2,m_h^2). 
\end{align}
Only for the top Yukawa coupling, the momentum assignment given in Eq.~(\ref{kappa_f}) is not kinematically allowed, so that
we assign the external momenta by $p_1^2=m_t^2$, $p_2^2=(m_t+m_h)^2$ and $q^2=m_h^2$ so as to be the on-shell top-quark and
the Higgs boson, which is related to the process; $e^+e^-\to t^* \bar{t}\to t\bar{t} h$. 

\begin{figure}[t]
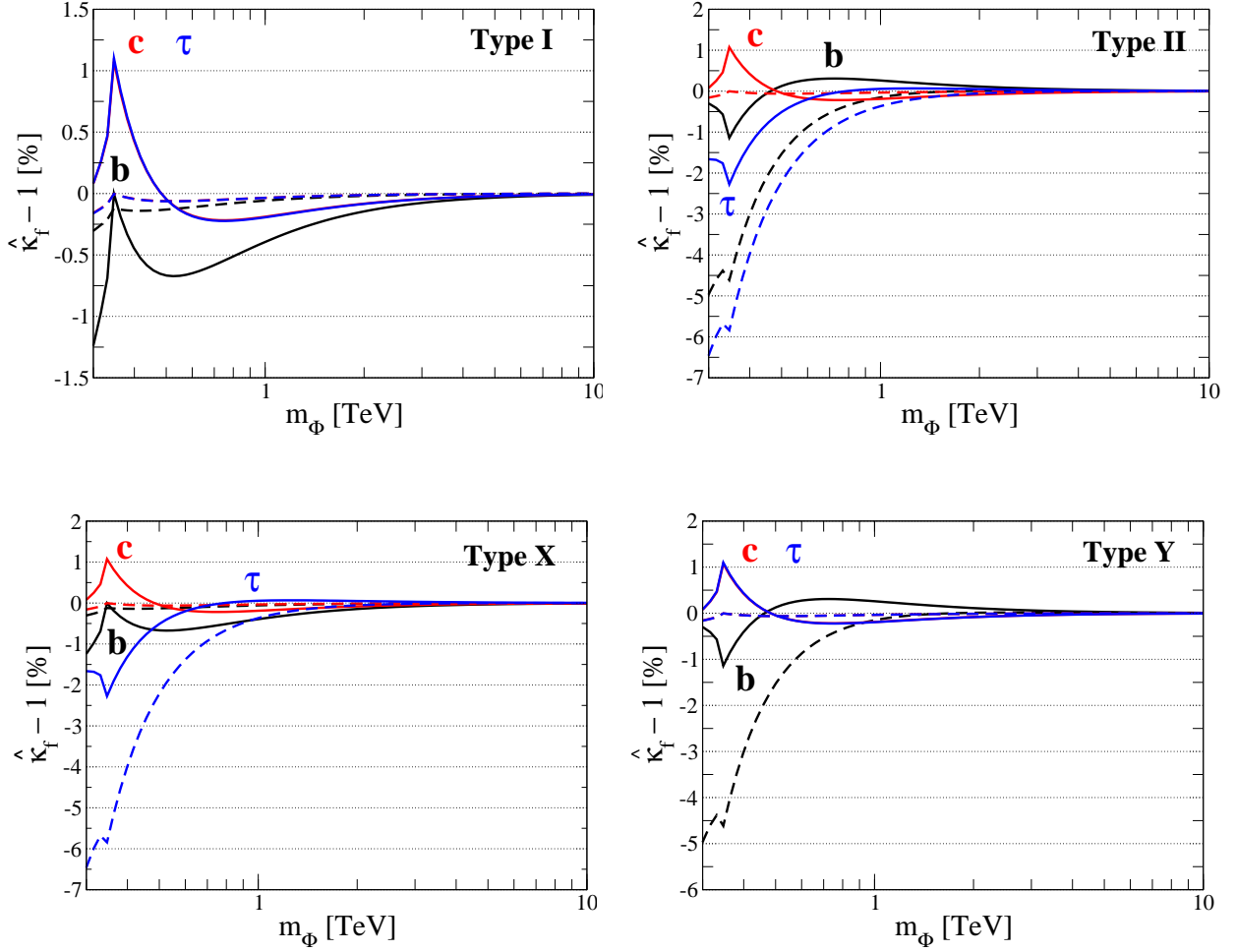

\begin{center}
\includegraphics[scale=0.33]{Dec_1.eps}\hspace{3mm}
\includegraphics[scale=0.33]{Dec_2.eps}\\\vspace{10mm}
\includegraphics[scale=0.33]{Dec_3.eps}\hspace{3mm}
\includegraphics[scale=0.33]{Dec_4.eps}\\\vspace{10mm}
\caption{Deviations in the renormalized Yukawa couplings for $b$, $\tau$ and $c$ 
as a function of $m_{\Phi}~(=m_{H^+}=m_A=m_H)$ in the case of $\sin(\beta-\alpha)=1$
in the Type-I (upper-left), Type-II (upper-right), Type-X (lower-left) and Type-Y (lower-right) THDMs.
The value of $M^2$ is taken so as to keep the relation (300 GeV)$^2=m_{\Phi}^2-M^2$.  
The solid and dashed curves are the results with $\tan\beta=1$ and $\tan\beta=3$, respectively. }
\label{fig1}
\end{center}
\end{figure}

In Fig.~\ref{fig1}, we first show the decoupling behavior of the one-loop contributions to the $hf\bar{f}$ couplings. 
As an example to see the decoupling, we only show the case with $\lambda_i v^2 = (300$ GeV)$^2$ (see Eq.~(\ref{mphi})) 
which corresponds to the case where the value of $M^2$ is changed to keep the relation (300 GeV)$^2=m_{\Phi}^2-M^2$. 
In this figure, 
the deviations in the renormalized Yukawa couplings; i.e., $\hat{\kappa}_f-1$
for $f=b$, $\tau$ and $c$ are shown as a function of $m_{\Phi}$ 
in the Type-I (upper-left), Type-II (upper-right), Type-X (lower-left) and Type-Y (lower-right) THDMs with $\sin(\beta-\alpha)=1$.
The solid and dashed curves are the results with $\tan\beta=1$ and $\tan\beta=3$, respectively.
In the large mass region, the value of $\hat{\kappa}_f-1$ asymptotically approaches to $0$ suggesting that 
the effects of the extra Higgs boson loops vanish. 
Thus, we can verify the reproduction of the SM prediction in the large mass limit. 
We note that the peak at around $m_{\Phi}=2m_t$ comes from the resonance of the top quark loop contribution to 
$\Pi_{AA}^{\text{1PI}}(p^2=m_A^2)$ which appears from the renormalization condition of $\delta\beta$.  

\begin{figure}[t]
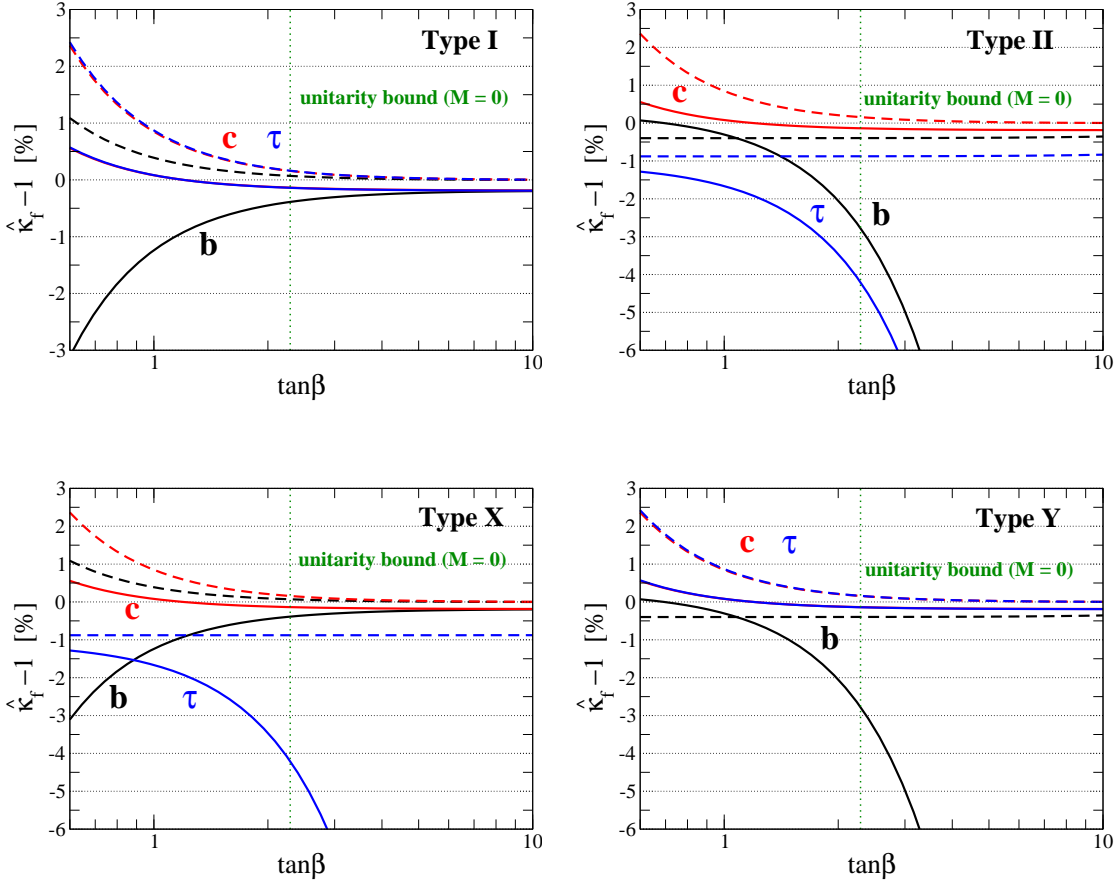

\begin{center}
\includegraphics[scale=0.3]{Tanb_1.eps}\hspace{3mm}
\includegraphics[scale=0.3]{Tanb_2.eps}\\\vspace{10mm}
\includegraphics[scale=0.3]{Tanb_3.eps}\hspace{3mm}
\includegraphics[scale=0.3]{Tanb_4.eps}
\caption{
Deviations in the renormalized Yukawa couplings for $b$, $\tau$ and $c$ 
as a function of $\tan\beta$ in the case of $\sin(\beta-\alpha)=1$
in the Type-I (upper-left), Type-II (upper-right), Type-X (lower-left) and Type-Y (lower-right) THDMs.
The extra Higgs boson masses $m_{\Phi}$ are taken to be 300 GeV in all the plots.   
The solid and dashed curves are the results with $M=0$ and $300$ GeV, respectively. 
For the case of $M=0$, the upper limit on $\tan\beta$ from the unitarity bound 
is denoted by the vertical dotted line (at around $\tan\beta\sim$ 2.3). 
}
\label{fig2}
\end{center}
\end{figure}

In Fig.~\ref{fig2}, we show the $\tan\beta$ dependence in $\hat{\kappa}_f$ for $f=b$, $\tau$ and $c$ in the 
Type-I (upper-left), Type-II (upper-right), Type-X (lower-left) and Type-Y (lower-right) THDMs with $\sin(\beta-\alpha)=1$.
We set the extra Higgs boson masses $m_{\Phi}$ to be 300 GeV and $M$ to be 0 (solid curves) and 300 GeV (dashed curves).   
In the case of $M=0$, $\tan\beta\gtrsim 2.3$ is excluded by the unitarity bound. 
In the Type-II THDM, the magnitude of $\hat{\kappa}_b$ and $\hat{\kappa}_\tau$ is increased 
as $\tan\beta$ is getting larger values 
because of the term proportional to $\lambda_{\Phi\Phi h}\lambda_{\Phi\Phi H}$ in Eq.~(\ref{limit1}).
Similar behavior can be seen in $\hat{\kappa}_\tau$ ($\hat{\kappa}_b$) in the Type-X (Type-Y) THDM.
In the Type-I THDM, such an enhancement does not appear because of the factor $\cot\beta$. 
We note that, although in Fig.~2 the results are shown for $0.6<\tan\beta<10$,
the case of $\tan\beta<1$ has been disfavored by the $B$ physics experiments such as 
$b\to s\gamma$ and the $B$-$\bar{B}$ mixing~\cite{Stal} in four types of Yukawa interactions.

\begin{figure}[t]
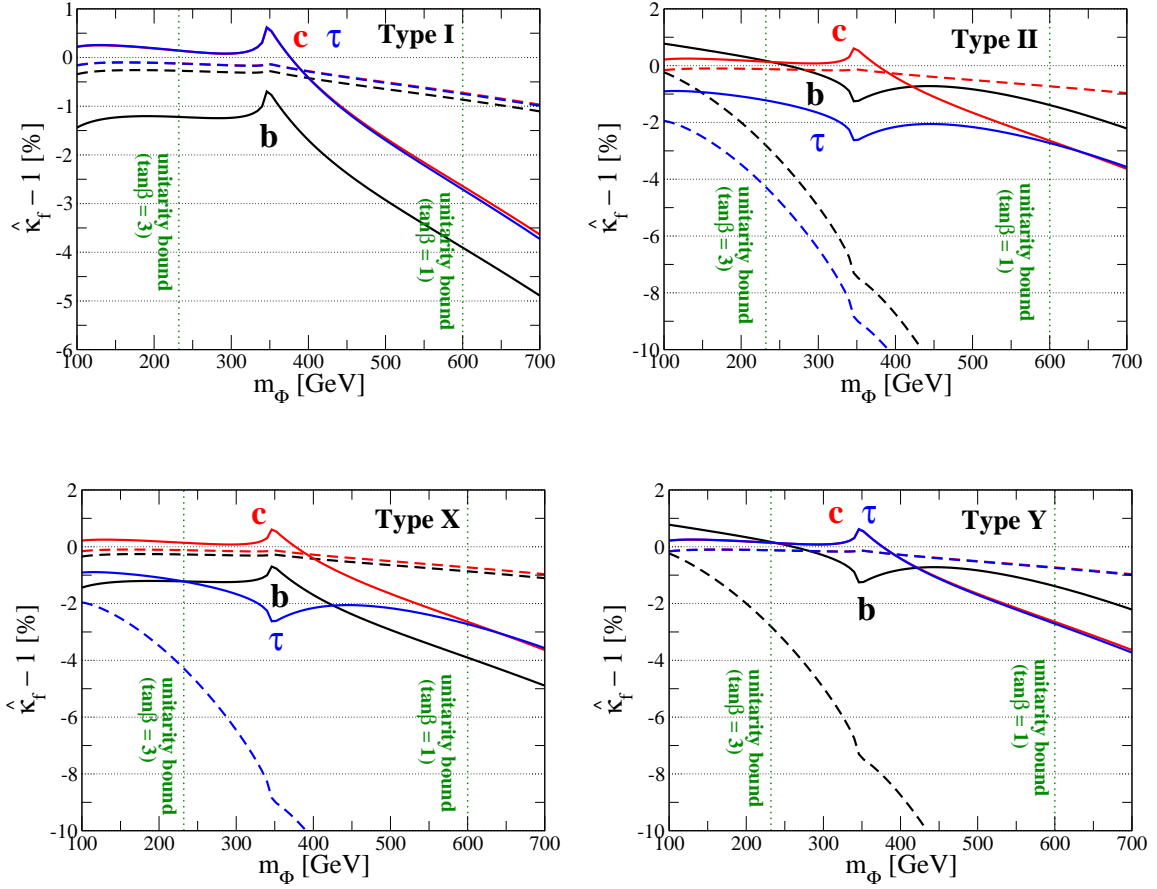

\begin{center}
\includegraphics[scale=0.3]{Nondec_1.eps}\hspace{3mm}
\includegraphics[scale=0.3]{Nondec_2.eps}\\\vspace{10mm}
\includegraphics[scale=0.3]{Nondec_3.eps}\hspace{3mm}
\includegraphics[scale=0.3]{Nondec_4.eps}
\caption{
Deviations in the renormalized Yukawa couplings for $b$, $\tau$ and $c$ 
as a function of $m_{\Phi}$ in the case of $\sin(\beta-\alpha)=1$
in the Type-I (upper-left), Type-II (upper-right), Type-X (lower-left) and Type-Y (lower-right) THDMs.
We take $M=0$ in all the plots. 
The solid and dashed curves are the results with $\tan\beta=1$ and $3$ GeV, respectively. 
The upper limits of $m_\Phi$ are denoted by the vertical dotted lines 
(at around $m_{\Phi}\simeq 600$ and 230 GeV for $\tan\beta=$1 and 3, respectively) from the unitarity bound. 
}
\label{fig3}
\end{center}
\end{figure}

\begin{figure}[t]
\begin{center}
\includegraphics[width=100mm]{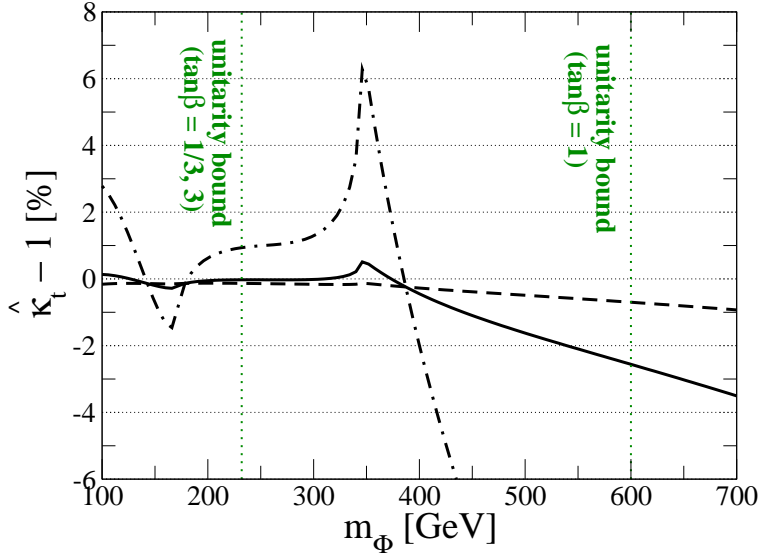}
\caption{
Deviations in the renormalized top Yukawa couplings 
as a function of $m_{\Phi}$ in the case of $\sin(\beta-\alpha)=1$ and $M=0$. 
The dash-dotted, solid and dashed curves are the results with $\tan\beta=1/3,~1$ and $3$, respectively. 
The upper limits of $m_\Phi$ are denoted by the vertical dotted lines 
at around $m_{\Phi}\simeq 600$ and (230) GeV for $\tan\beta=$1 and (3 and $1/3$), respectively from the unitarity bound. 
}
\label{fig4}
\end{center}
\end{figure}

\begin{figure}[!t]
\begin{center}
\vspace{-15mm}
\includegraphics[scale=0.3]{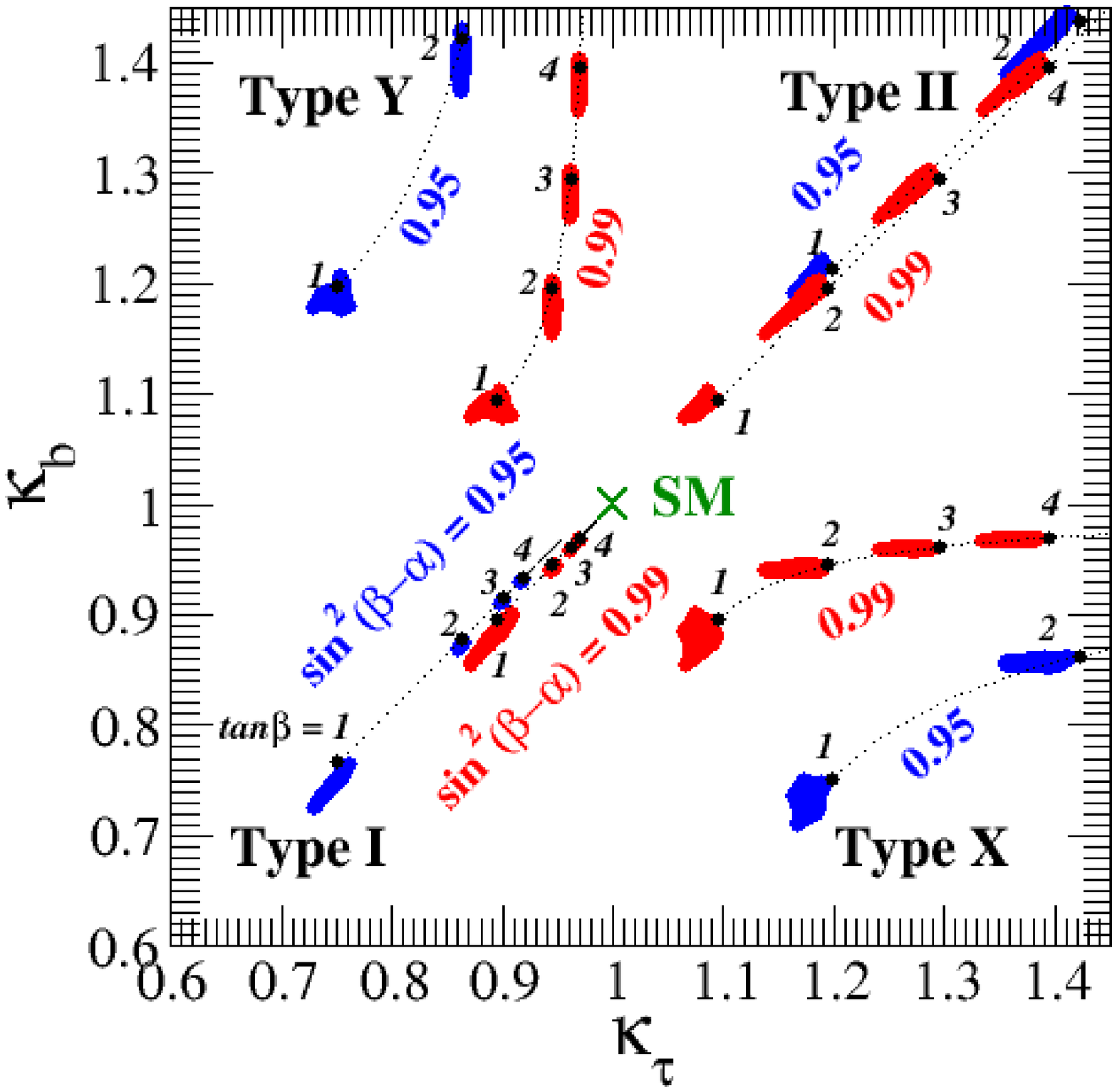}\hspace{-21mm}
\includegraphics[scale=0.3]{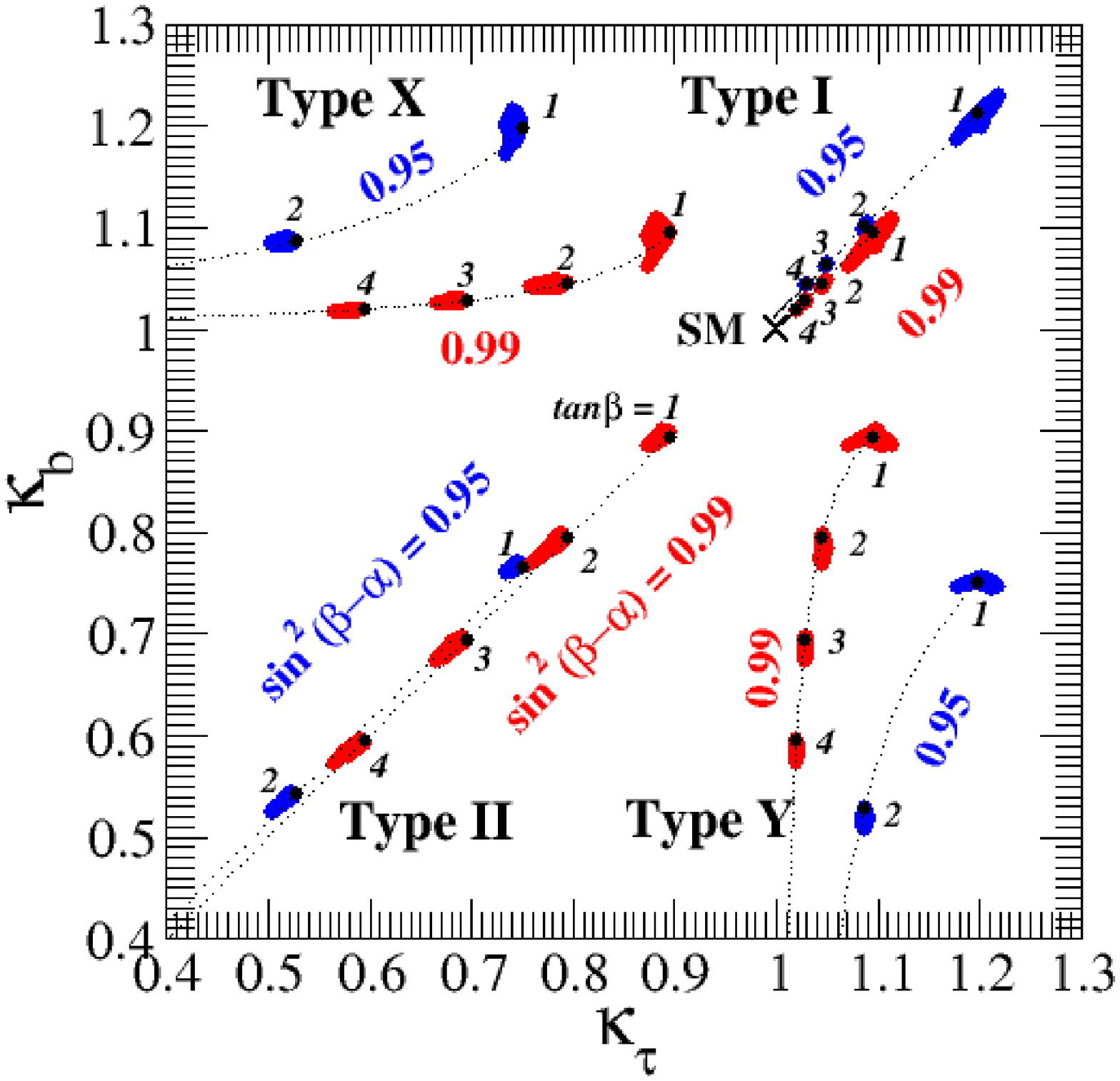}\\\vspace{-3mm}
\includegraphics[scale=0.3]{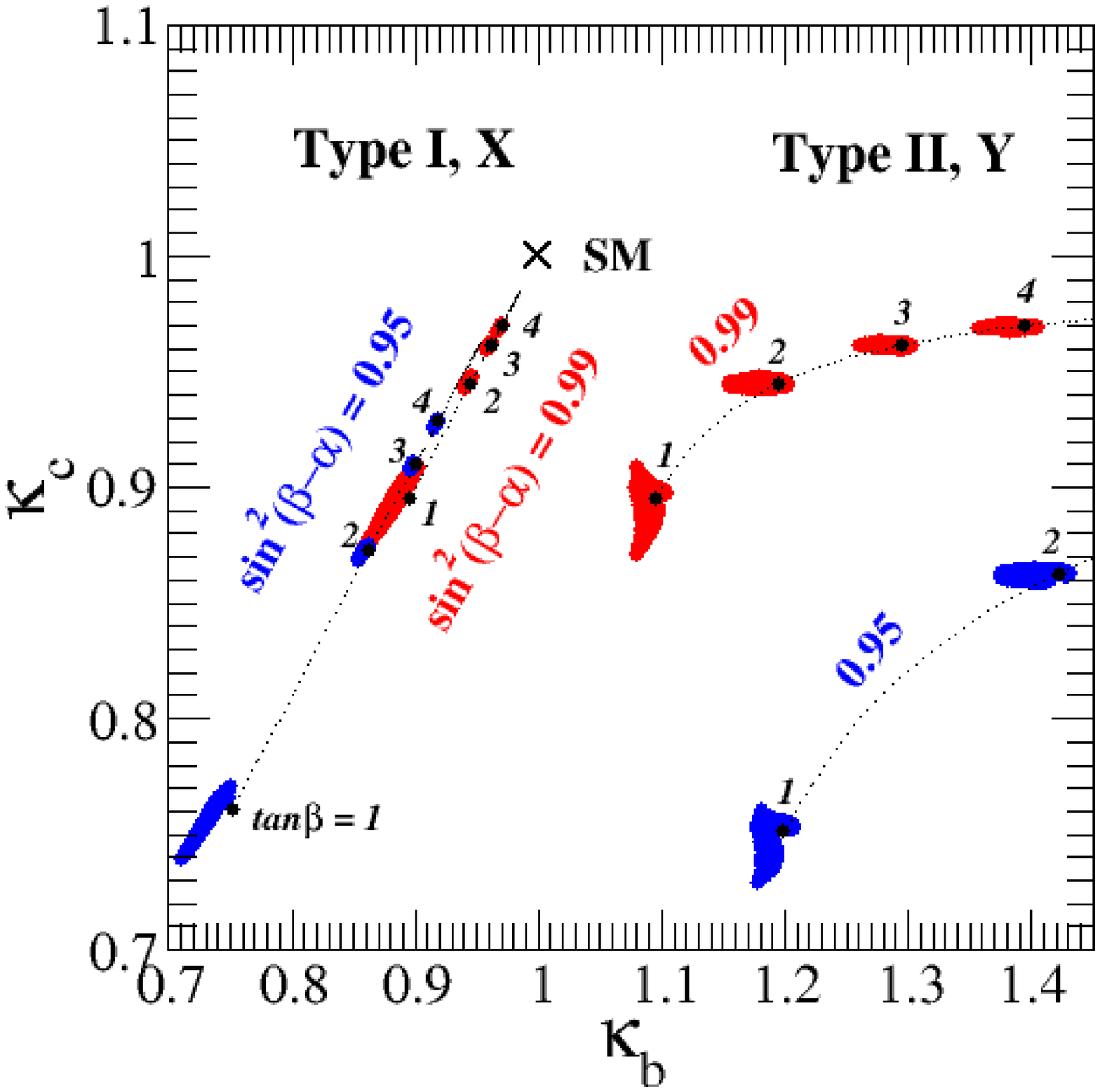}\hspace{-21mm}
\includegraphics[scale=0.3]{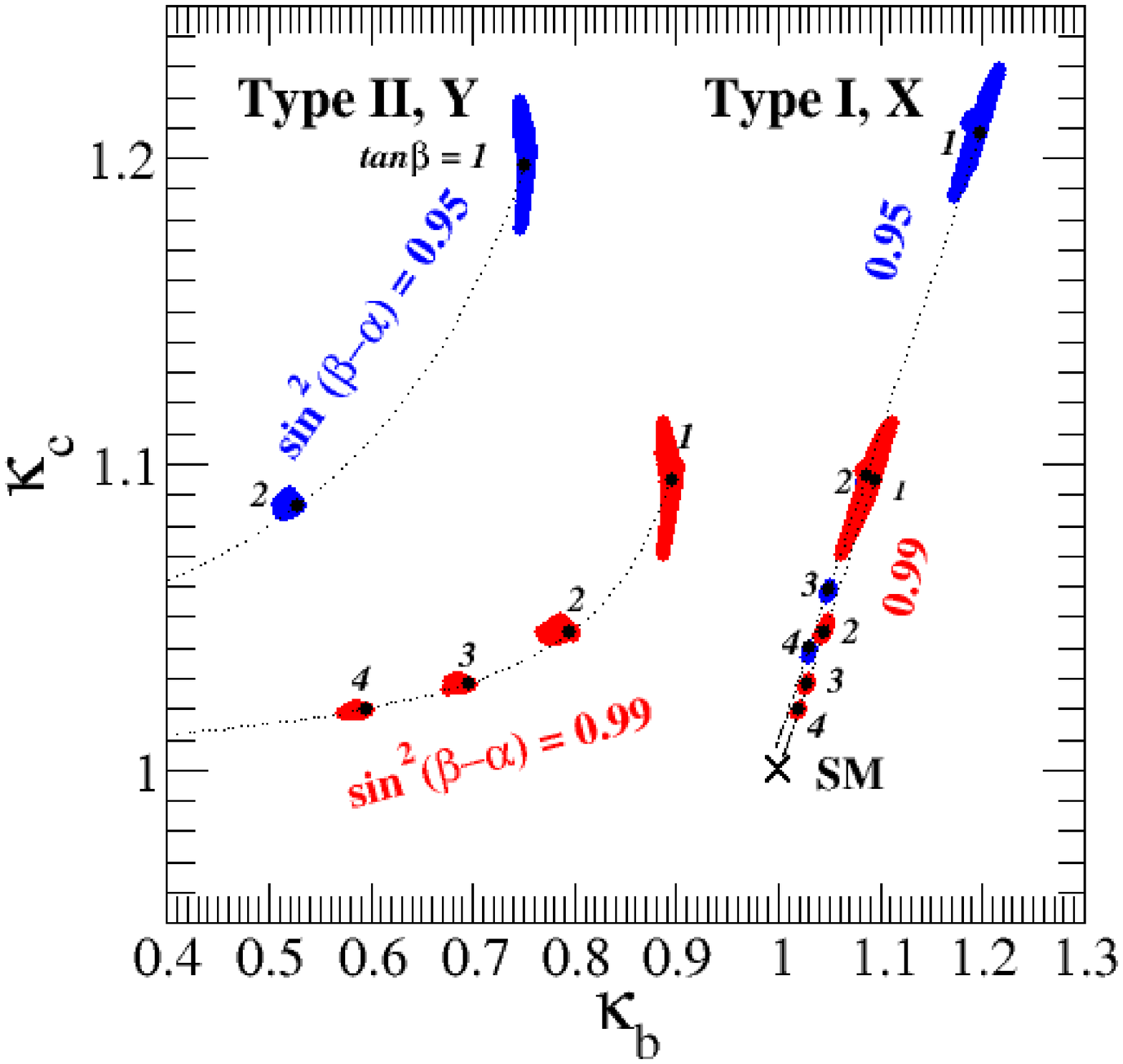}\\\vspace{-3mm}
\includegraphics[scale=0.3]{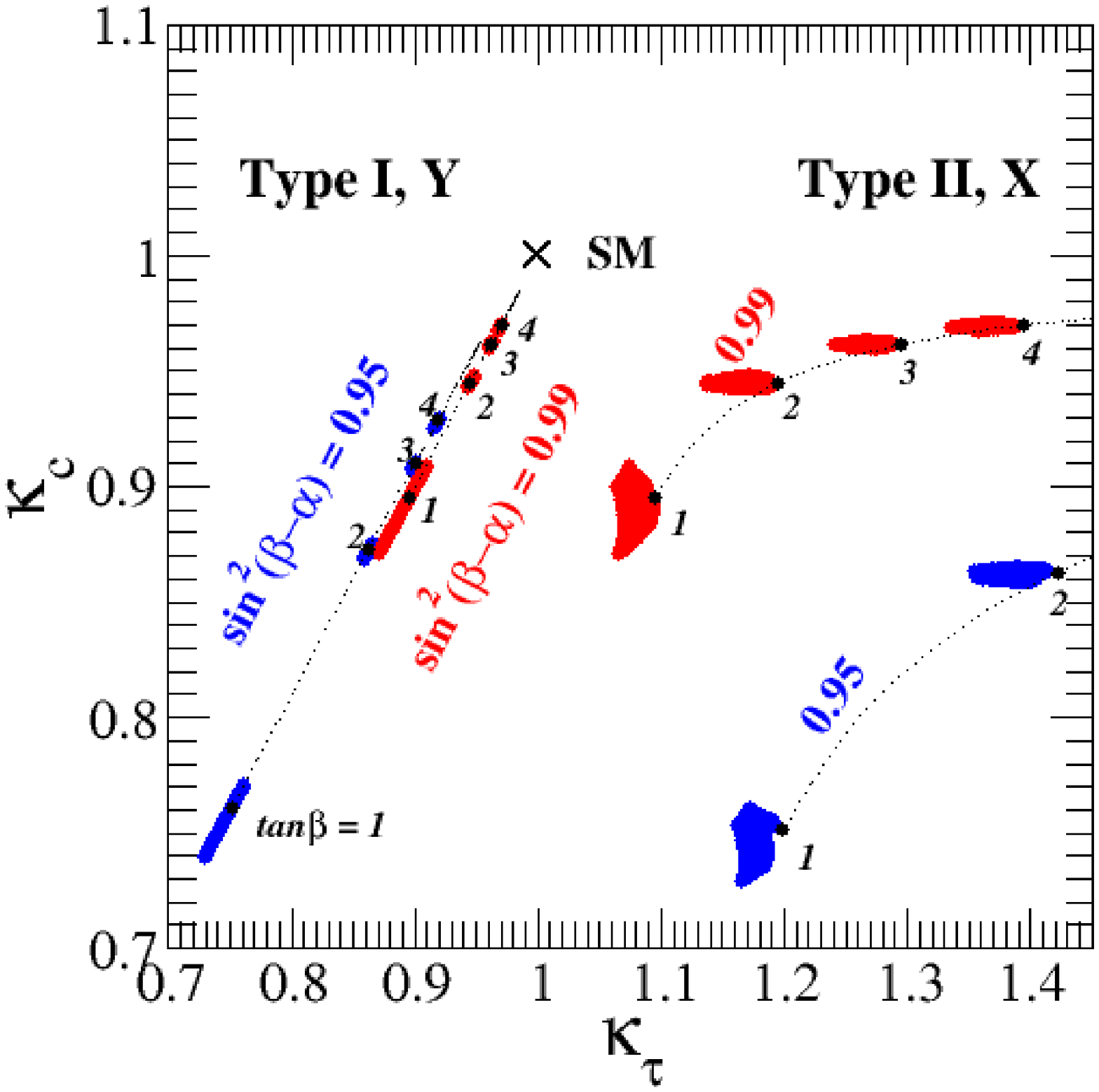}\hspace{-21mm}
\includegraphics[scale=0.3]{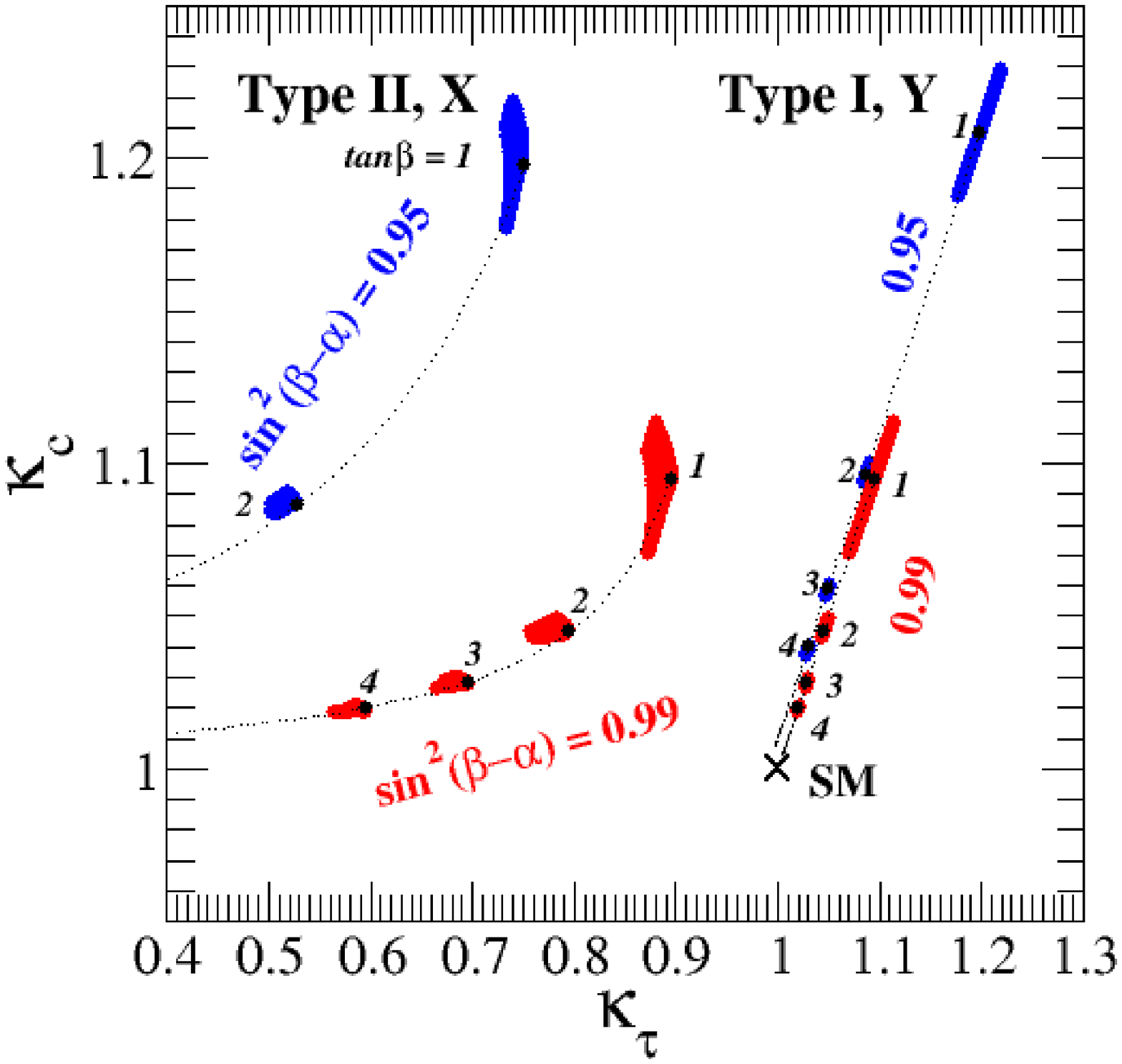}
\caption{Predictions of various scale factors on the $\kappa_\tau$ vs $\kappa_b$ (upper panels), 
$\kappa_b$ vs $\kappa_c$ (middle panels) and $\kappa_\tau$ vs $\kappa_c$ (bottom panels) planes in 
four types of Yukawa interactions. 
The left and right three figures show the cases with $\cos(\beta-\alpha)<0$ and $\cos(\beta-\alpha)>0$, respectively. 
Each black dot shows the tree level result with $\tan\beta$=1,~2,~3 and 4. 
One-loop corrected results are indicated by red for $\sin^2(\beta-\alpha)=0.99$ and 
blue for $\sin^2(\beta-\alpha)=0.95$ regions where 
$m_\Phi$ and $M$ are scanned over from 100 GeV to 1 TeV and 0 to $m_\Phi$, respectively. 
All the plots are allowed by the unitarity and vacuum stability bounds.  
}
\label{fig5}
\end{center}
\end{figure}

Next, we show the nondecoupling effect due to the extra Higgs boson loops to the $hf\bar{f}$ couplings. 
Such an effect can be extracted from Eqs.~(\ref{limit1}) and (\ref{limit2}) symbolically as; 
\begin{align}
\hat{\Gamma}_{hff}^{\text{THDM}}
\sim 
\hat{\Gamma}_{hff}^{\text{SM}}
+\frac{1}{16\pi^2}\frac{m_f}{v}\frac{m_\Phi^2}{v^2}\left(1-\frac{M^2}{m_\Phi^2}\right)^2. \label{nondec}
\end{align}
From the above expression, it is clarified that there appears the quadratic dependence of $m_\Phi$. 
Such a quadratic dependence vanishes when $M\simeq m_\Phi$. 

In Fig.~\ref{fig3}, the $m_\Phi$ dependence in $\hat{\kappa}_f$ for $f=b$, $\tau$ and $c$ is shown 
in the Type-I (upper-left), Type-II (upper-right), Type-X (lower-left) and Type-Y (lower-right) THDMs with $\sin(\beta-\alpha)=1$.
The solid and dashed curves are the results with $\tan\beta=1$ and $3$, respectively. 
We here take $M^2=0$ to see the nondecoupling effect in all the plots.\footnote{
if we take negative values for $M^2$, larger nondecoupling effects can be obtained as compared to the case with $M^2=0$. 
However, too large negative values for $M^2$ are easily excluded by perturbative unitarity. } 
The maximal value of $m_\Phi$ is constrained by the unitarity bound; 
i.e., $m_{\Phi}\gtrsim 600$ GeV (230 GeV) is excluded in the case with $\tan\beta=1$ (3) as shown by the vertical dotted lines. 
In the case of $\tan\beta=1$, the maximal allowed deviations in $\hat{\kappa}_f$ are about from $-2\%$ to $-5\%$ depending on the 
types of Yukawa interactions.

In the above discussions, we consider Yukawa couplings for the bottom quark, charm quark and tau lepton.  
Let us discuss the top Yukawa coupling. 
As already mentioned in the beginning of this section, 
only the top Yukawa coupling is treated as different way from the other fermions; namely, 
$\hat{\kappa}_t$ is defined by 
\begin{align}
\hat{\kappa}_t \equiv
\frac{
\hat{\Gamma}_{htt}(m_t^2,(m_t+m_h)^2,m_h^2)_{\text{THDM}}
}{\hat{\Gamma}_{htt}(m_t^2,(m_t+m_h)^2,m_h^2)_{\text{SM}}}.
\end{align}
In Fig.~\ref{fig4}, deviations in the renormalized top Yukawa coupling $\hat{\kappa}_t-1$
are shown as a function of $m_{\Phi}$ in the case of $\sin(\beta-\alpha)=1$ and $M=0$. 
The value of $\tan\beta$ is fixed by 1/3 (dashe-dotted)\footnote{
As we already mentioned before, the case of $\tan\beta=1/3$ has been excluded by the $B$ physics data. 
Nevertheless, we show the case with $\tan\beta=1/3$ just for the reference. 
}, 1 (solid curve) and 3 (dashed curve). 
The difference in $\hat{\kappa}_t$ among the types of Yukawa interactions can be neglected similar to $\hat{\kappa}_c$. 
The height of the peak at around $m_\Phi=2m_t$ depends on $\cot^2\beta$, so that we can see the large peak in the case of $\tan\beta=1/3$. 
The maximal allowed amount for the deviation in the top Yukawa coupling is about $+4~\%$, 
$-6~\%$ and $-1~\%$ for the cases with 
$\tan\beta=1/3,~1$ and 3, respectively.

Finally, we show the one-loop results for the Yukawa couplings in the planes of fermion scale factors.
In Fig.~\ref{fig5}, predictions of various scale factors are shown 
on the $\kappa_\tau$ vs $\kappa_b$ (upper panels), 
$\kappa_b$ vs $\kappa_c$ (middle panels) and $\kappa_\tau$ vs $\kappa_c$ (bottom panels) planes. 
When we consider the case with $\sin(\beta-\alpha)\neq 1$, the sign dependence of $\cos(\beta-\alpha)$ to 
$\hat{\kappa}_f$ is also important as we can see Eqs.~(\ref{ku}), (\ref{kd}) and (\ref{ke}). 
Thus, we show the both cases with $\cos(\beta-\alpha)<0$ (left panels) and $\cos(\beta-\alpha)>0$ (right panels). 
The value of $\tan\beta$ is discretely taken as $\tan\beta$=1,~2,~3 and 4. 
The tree level predictions are indicated by the black dots, while 
the one-loop corrected results are shown by the red for $\sin^2(\beta-\alpha)=0.99$ and 
blue for $\sin^2(\beta-\alpha)=0.95$ regions where 
the values of $m_\Phi$ and $M$ are scanned over from 100 GeV to 1 TeV and 0 to $m_\Phi$, respectively. 
All the plots are allowed by the unitarity and vacuum stability bounds. 

The tree level behaviors on $\kappa$-$\kappa$ panels 
can be understood by looking at the expressions given in Eqs.~(\ref{ku}), (\ref{kd}) and (\ref{ke}). 
In the middle and bottom panels, predictions in two of four THDMs are degenerate at the tree level; e.g., 
results in the Type-I and Type-X THDMs are the same on the $\kappa_b$-$\kappa_c$ panel. 
This is because the same Higgs doublet field couples to corresponding fermions, which can be understood 
more clearly by looking at the expression given in Eqs.~(\ref{ku}), (\ref{kd}) and (\ref{ke}). 
On the other hand, 
in the $\kappa_\tau$-$\kappa_b$ plane, predictions in all the four types are located in different areas with each other. 
Thus, all the types of THDMs give different predictions by looking at all three combinations of $\kappa$-$\kappa$ planes.

Even when we take into account the one-loop corrections to the Yukawa couplings,   
this behavior; i.e., predictions are well separated among the four types of THDMs, does not so change as we see the 
red and blue colored regions. 
Therefore, we conclude that 
all the THDMs can be distinguished from each other by measuring the charm, bottom and tau Yukawa couplings precisely when 
the gauge couplings $hVV$ are deviated from the SM prediction with $\mathcal{O}(1)$\%. 

We here comment on the $hVV$ couplings in the THDMs. 
Although the tree level deviations in the $hVV$ couplings are described by the factor $\sin(\beta-\alpha)$, 
these values can be modified at the one-loop level.  
In Ref.~\cite{KOSY}, the one-loop corrected $hZZ$ vertex has been calculated in the softly-broken $Z_2$ symmetric THDM.  
It has been found that 
for the fixed value of $\sin(\beta-\alpha)$, 
the one-loop corrections to the $hZZ$ vertex are less than $1\%$ even taking the maximal nondecoupling case.

Before the conclusions, we mention about the expected accuracy for the various Higgs boson couplings 
measured at future colliders such as the LHC with the 14 TeV run and the ILC. 
According to the ILC Technical Design Report~\cite{ILC_TDR,ILC_white}, 
the $hVV$ couplings are expected to be measured with about 4$\%$ accuracy at the LHC with 300~fb$^{-1}$. 
The accuracy for the $ht\bar{t}$, $hb\bar{b}$ and $h\tau\tau$ couplings are supposed to be about 16\%, 14\% and 11\%, respectively. 
At the ILC250 (ILC500) where the collision energy and the integrated luminosity are 250 GeV (500 GeV) and 
250 fb$^{-1}$ (500 fb$^{-1}$) combining with the results assuming 300~fb$^{-1}$ at the LHC, 
the $hWW$ and $hZZ$ couplings are expected to be measured by about 1.9\% (0.2\%) and about 0.4\% (0.3\%), respectively. 
The $hc\bar{c}$, $hb\bar{b}$ and $h\tau\tau$ couplings are supposed to be measured by
about 5.1\% (2.6\%), 2.8\% (1.0\%) and 3.3\% (1.8\%) at the ILC250 (ILC500).  
For the $ht\bar{t}$ coupling, it will be measured with 12.0\% and 9.6\% accuracy at the ILC250 and ILC500, respectively. 
Therefore, if $\mathcal{O}(1)\%$ deviations in the $hVV$ couplings from the SM values 
are established at the ILC250, we can compare the predictions of $\hat{\kappa}_f$ 
to the corresponding measured values at the ILC500, which are typically measured by $\mathcal{O}(1)\%$.
We can then discriminate the types of Yukawa interactions in the THDM.  

\section{Conclusions}

We have evaluated radiative corrections to the $hf\bar{f}$ couplings in the THDMs 
with the softly-broken $Z_2$ symmetry.
We have found that one-loop contributions of extra Higgs bosons can 
modify the $hf\bar{f}$ couplings to be maximally about $5\%$ 
under the constraint from perturbative unitarity and vacuum stability.
The results indicate that 
the pattern of tree-level deviations by the mixing effect 
in each type of Yukawa couplings from the SM predictions does not change 
even including radiative corrections.  
Moreover, when the gauge couplings $hVV$ will be found to be 
slightly (with a percent level) differ from the SM predictions, 
the $hf\bar{f}$ couplings also deviate but more largely. 
In this case, by comparing the predictions with precisely measured $hf\bar{f}$ and  
$hVV$ couplings at the ILC, we can determine the type of Yukawa couplings 
and also can obtain information on the extra Higgs bosons, even when 
they are not found directly.
\\\\
\noindent
$Acknowledgments$

S.K. was supported in part by Grant-in-Aid for Scientific Research, Japan
Society for the Promotion of Science (JSPS), Nos. 22244031 and 24340046, and The Ministry of Education, Culture, 
Sports, Science and Technology (MEXT), No. 23104006.
M.K. was supported in part by JSPS, No. K2510031.
K.Y. was supported in part by the National Science Council of R.O.C. under Grant No. NSC-101-2811-M-008-014.


\end{document}